\documentclass[epj]{webofc}
\usepackage[varg]{txfonts}   
\usepackage{bm}
\usepackage{amssymb}
\usepackage{amsmath}
\usepackage{enumerate}
\usepackage{epsf}

\wocname{EPJ Web of Conferences}
\def\S{\mathcal{S}}

\def\eRM{{\mathrm e}}
\def\dRM{{\mathrm d}}

\def\mx{{\bm x}}
\def\mv{{\bm v}}

\def\mk{{\bm k}}
\def\mf{{\bm f}}
\def\mp{{\bm p}}

\def\eps{\varepsilon}
\def\boldnabla{{\bm \nabla}}
\allowdisplaybreaks
\allowbreak
\newcommand{\fp}[1]{FP$ {\textrm{#1}}$}

\begin{document}
\selectlanguage{english}
\title{Renormalization group analysis of a turbulent compressible fluid near $d=4$: Crossover 
between local and non-local scaling regimes.}
%
%

\author{N. V. Antonov\inst{1}\fnsep\thanks{\email{n.antonov@spbu.ru}} \and
        N. M. Gulitskiy\inst{1}\fnsep\thanks{\email{n.gulitskiy@spbu.ru}} \and
        M. M. Kostenko\inst{1}\fnsep\thanks{\email{kontramot@mail.ru}} \and
        T. Lu\v{c}ivjansk\'y\inst{2}\fnsep\thanks{\email{tomas.lucivjansky@uni-due.de}}
        }

\institute{
Department of Theoretical Physics, Faculty of Physics,  
Saint-Petersburg State University, 7/9~Universitetskaya nab., St. Petersburg, 199034 Russia
\and
Faculty of Sciences, P.J. \v{S}af\'arik
University, Moyzesova 16, 040 01 Ko\v{s}ice, Slovakia,\\ and
Fakult\"at f\"ur Physik, Universit\"at Duisburg-Essen, D-47048 Duisburg, Germany
}

\abstract{
We study scaling properties of the model of fully developed turbulence for a
compressible fluid, based on the stochastic Navier-Stokes equation, by means of the field
theoretic renormalization group (RG). 
The scaling properties in this approach are related to fixed points of the RG equation.
Here we study a possible existence of other scaling regimes and an opportunity
of a crossover between them. This may take place in some other space dimensions, 
particularly at $d = 4$. A new regime may there arise and then by continuity moves into
$d = 3$. 
Our calculations have shown that there really exists an additional fixed point, that may
govern scaling behaviour.
}

\maketitle
{\section{Introduction}\label{sec:intro}}
A majority of works on fully developed turbulence is concerned with an incompressible fluid. The 
renormalization
group approach to such problems has been successful in verifying Kolmogorov scaling and provides
an efficient tool for a calculation of universal quantities. However, a similar  treatment has
been only scarcely applied to compressible fluids. In this paper we present an application 
of the field
theoretic renormalization group (RG) onto the scaling regimes of a compressible fluid, whose
behavior is governed by a proper generalization of stochastic Navier Stokes equation~\cite{ANU97}. 
Similar models of compressible fluid were considered in~\cite{LM,St,MTY}. In~\cite{LM} the phenomenological corrections to the Kolmogorov spectrum were verified in the framework of the skeleton equatios for consistency, while the model, considered in~\cite{St}, appears to be in fact unrenormalizible. All these papers shows us a necessity of the further investigations of compressibility.

Following~\cite{HN96}, we employ 
double expansion scheme. Here the formal expansion 
parameters are $y$, which describes the
scaling behavior of a random force, and $\eps=4-d$, i.e., a deviation from the dimension of space $d=4$.

{\section{Description of the model} \label{sec:desc}}
The Navier-Stokes equation for a compressible fluid can be written in the following form:
\begin{equation}
   \rho[\partial_t \mv + (\mv\cdot\boldnabla)\mv] = \nu_0[\boldnabla^2 \mv - \boldnabla (\boldnabla\cdot\mv)]
   +\mu_0 \boldnabla (\boldnabla\cdot\mv) - \boldnabla p + \mf,
   \label{eq:compNS1}
\end{equation}
where $\rho$ is the fluid density, $\mv$ is the velocity field,
$\partial_t$ is a time derivative $\partial/\partial t, \boldnabla^2$ is the Laplace operator, $\nu_0$ and $\mu_0$ are molecular
viscosity coefficients, $p$ is pressure field,  and $\mf$ is an external field per unit mass. 
 The model must be augmented by two additional equations, namely a continuity equation and an equation of state between 
 deviations $\delta p$ and $\delta \rho$ from the equilibrium values. They read
\begin{subequations}
\begin{align}
  \partial_t \rho + \boldnabla\cdot (\rho\mv) &= 0;\\
  \delta p = c_0^2 \delta\rho.&
   \end{align}
  \label{eq:continuity}
\end{subequations} 
In order to obtain the renormalizable field theoretic model  expression~(\ref{eq:compNS1}) is
divided by $\rho$, and fluctuations in viscous terms are neglected~\cite{VN96}. Further, by using the continuity
equation and the equation of state~(\ref{eq:continuity}), the problem can be recasted in terms of two coupled equations:
\begin{subequations}
\begin{align}
   (\partial_t  + \mv\cdot\boldnabla)\mv  &= \nu_0[\boldnabla^2 \mv - \boldnabla (\boldnabla\cdot\mv)]
   +\mu_0 \boldnabla (\boldnabla\cdot\mv) - \boldnabla \phi + \mf;  \\
   (\partial_t + \mv\cdot\boldnabla)\phi  &= -c_0^2 \ (\boldnabla\cdot\mv).
   \end{align}
\label{eq:model}
\end{subequations} 
Here $\phi$ is related to the density fluctuations via the relation $\phi = c_0^2 \ln (\rho/\overline{\rho})$.
Parameter $c_0$ is an adiabatic speed of sound, $\overline{\rho}$ denotes the mean value of $\rho$.

The turbulence is modeled by an external force -- it is assumed to be a random variable, which
should mimic the input of the energy into the system from the outer large scale $L$. Its precise
form is believed to be unimportant and is usually considered to be a random Gaussian variable with zero mean
and correlator
\begin{subequations}
\begin{align}
  \langle f_i(t,\mx) f_j(t',\mx') &= \frac{\delta(t-t')}{(2\pi)^d} \int_{k>m} \dRM^d\mk\mbox{ } D_{ij}(\mk)
  \eRM^{i\mk\cdot(\mx-\mx')}, \quad \text{where} \\
  D_{ij}(\mk) &= g_{10} \nu_0^3 k^{4-d-y} \biggl\{  P_{ij}(\mk) + \alpha Q_{ij}(\mk) \biggl\}.
   \end{align}
  \label{eq:correl}
\end{subequations} 
Here $d$ is the space dimension, $P_{ij}(\mk) = \delta_{ij}-k_ik_j/k^2$ and $Q_{ij}(\mk)=k_ik_j/k^2$ are the transverse 
and longitudinal
projectors, $k=|\mk|$, a parameter $m=L^{-1}$ provides an infrared (IR) 
cutoff, amplitude $\alpha$ is a free parameter, an exponent $y$
 plays a role of a formally small expansion parameter, and $g_{10}$ is a coupling 
 constant; Dirac delta function ensures Galilean 
 invariance~\cite{turbo}. 

{\section{Field theoretic formulation of the model} \label{sec:model}}
According to the general theorem~\cite{Vasiliev,Tauber}, the stochastic problem is equivalent to the field theoretic
model with a doubled set of fields $\tilde{\psi},\psi$ and de Dominicis-Janssen action functional, written
in a compact form as
\begin{align}
  \S(\varphi) & = \frac{v_i' D_{ik}^f v_k'}{2} 
  +v_i' \biggl\{
  -\partial_t v_i - v_j\partial_j v_i + \nu_0[\delta_{ik}\partial^2 - \partial_i \partial_k]v_k
  +u_0 \nu_0 \partial_i \partial_k v_k - \partial_i \phi
  \biggl\} \nonumber \\  &
  +\phi'[-\partial_t \phi + v_j\partial_j \phi + v_0 \nu_0 \partial^2 \phi - c_o^2 (\partial_i v_i)].
  \label{eq:action} 
\end{align}
Here we have employed a condensed notation, in which integrals over the spatial variable 
${\bf x}$ and the time variable $t$, as well as summation over repeated indices, are implicitly assumed. 
The action~(\ref{eq:action}) is amenable to the standard methods
of the quantum field theory, such as the Feynman diagrammatic technique and 
the renormalization group procedure.

\begin{figure}[h!]
   \centering
   \begin{tabular}{c c}
     \includegraphics[width=6.1cm]{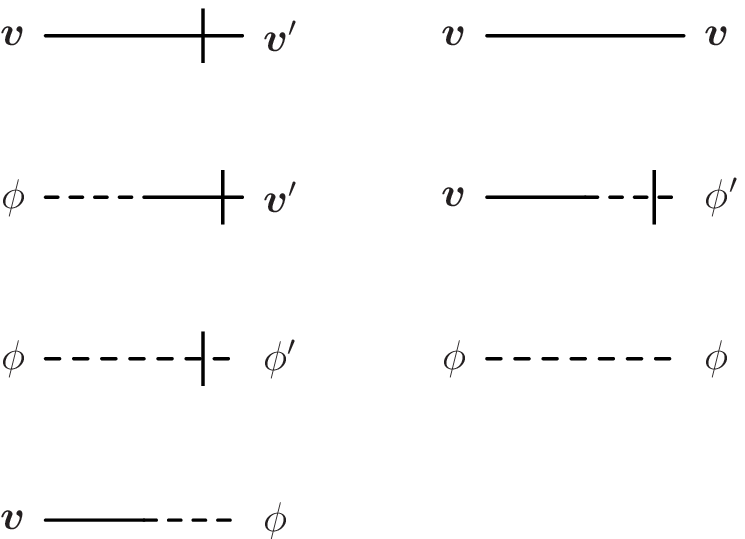}
     &
     \includegraphics[width=6.1cm]{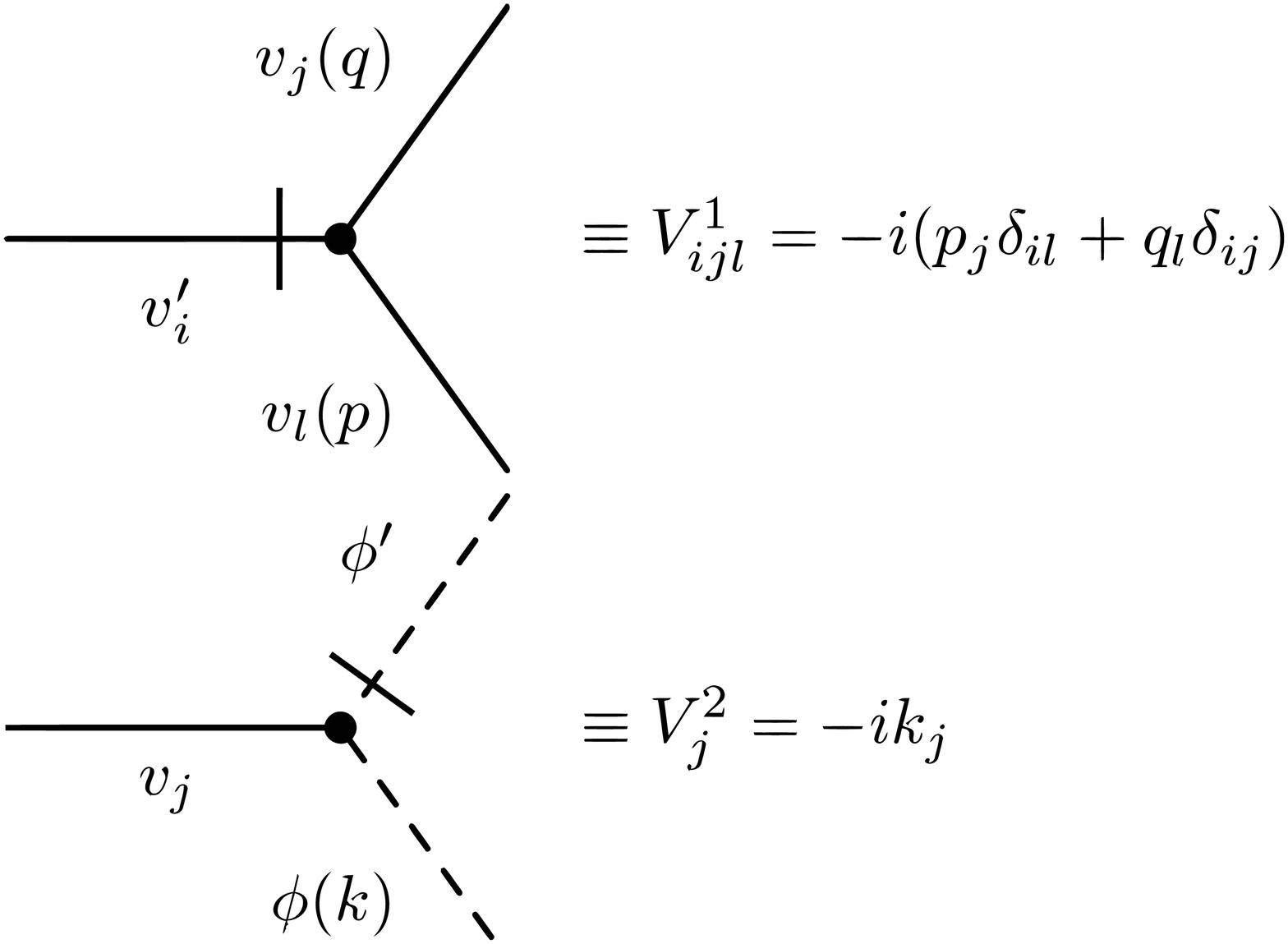}  
      \end{tabular}
   \caption{Graphical representation of the bare propagators and interaction vertices in the model~(\ref{eq:action})}
   \label{fig:prop_vertex}
\end{figure}

In a standard approach, if we apply quantum field methods to the stochastic differential equations, the space
dimension $d$ plays
a passive role and an actual perturbative parameter is $y$; for more details see the monographs~\cite{turbo,Vasiliev}. 
Our approach closely follows the analysis of the 
incompressible Navier-Stokes equation near
space dimension $d=2$ (see~\cite{HN96,AHKV03,AHKV05,AHH10}). In this case three  additional divergences appear in the Green's function $v'v'$. They can
be absorbed by a suitable local counterterm $v'_i\boldnabla^2 v'_i$, and a regular
expansion in both $y$ and $\eps'=d-2$ was constructed.
 Up to now the present model~(\ref{eq:action}) has been 
investigated at the fixed space dimension $d=3$, for which the action 
(\ref{eq:action}) contains all terms that can be
 generated during the renormalization procedure~\cite{ANS95,ANU97,AK14,AK15}. However, using the dimensional
analysis it can be shown that at $d=4$ there appears an additional divergence, also
in the Green's function
$v'v'$. Therefore, to keep the model renormalizable at $d=4$ the kernel function in~(\ref{eq:correl}) has to be generalized to the following form:
\begin{equation}  
  D_{ij}(\mk) \rightarrow g_{10} \nu_0^3 k^{4-d-y} \biggl\{
  P_{ij}(\mk) + \alpha Q_{ij}(\mk)
  \biggl\} 
  + g_{20} \nu_0^3 \delta_{ij},
  \label{eq:correl2}
\end{equation}
where the new term on the right hand side absorbs divergent contributions from $v'v'$. In contrast to
\cite{HN96} no momentum dependence is needed.
{\section{Feynman diagrammatic technique} \label{sec:technique}}
The perturbation theory of the model can be expressed in 
 the standard Feynman diagrammatic expansion~\cite{Zinn,Vasiliev}.
 Bare propagators are read off from the inverse matrix of the Gaussian (free) part of the action functional, while
the nonlinear part of the differential equation defines the interaction vertices.
Their graphical representation
is depicted in Fig.~\ref{fig:prop_vertex}. Explicit expressions of propagators in frequency-momentum representation
can be found, e.g., in~\cite{ANU97}, and they are right for actual calculations.
\newpage

The ultraviolet (UV) renormalizability is very efficiently revealed by an analysis
of the 1-irreducible Green's functions. Corresponding generating functional can be written in the form
\begin{equation}
  \Gamma(\varphi) =  \S(\varphi) + \widetilde{\Gamma}(\varphi),
  \label{eq:expansion}
\end{equation}
where for the functional arguments we have used the same symbols
$\varphi=\left\{\mv,\mv',\phi,\phi' \right\}$ as for the corresponding random fields~\cite{Vasiliev};
${\S}(\varphi)$ is the action functional~(\ref{eq:action}) and
$\widetilde{\Gamma}(\varphi)$ is the sum of all the 1-irreducible diagrams with loops~\cite{Vasiliev}.
As it has been shown in~\cite{ANU97} and discussed in~\cite{AK14}, the model~(\ref{eq:action})~--~(\ref{eq:correl2}) is
invariant with respect to the Galilean symmetry, which results to the UV finitness of the two 
Green's functions: $v_i\partial_t v_i$ and $v'_i (v_j\partial_j)v_i$.
We have carried on the perturbative analysis in the one-loop order, consequently the expressions for the 1-irreducible Green's functions, which requires UV renormalization, can be formally written 
in the following way:
\begin{align}
   \Gamma_{v'v} & = i\omega -
 (\delta_{ij}p^2 - p_i p_j)Z_1 \nu - p_i p_j Z_2 u\nu +  \raisebox{-1.ex}{ \includegraphics[width=2.5truecm]{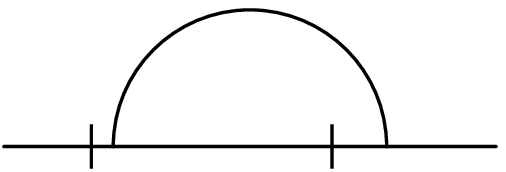}},
 \label{eq:vsv}\\
     \Gamma_{\phi \phi'} & = i\omega - p^2 Z_3 v\nu + \raisebox{-1ex}{ \includegraphics[width=2.5truecm]{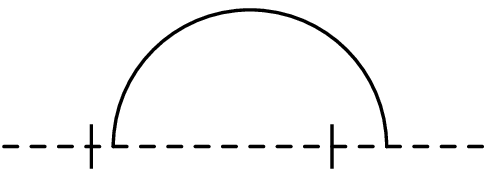}},
    \label{eq:phiphis}\\
    \Gamma_{v'\phi} & = -iZ_4 p_i +\raisebox{-1ex}{ \includegraphics[width=2.truecm]{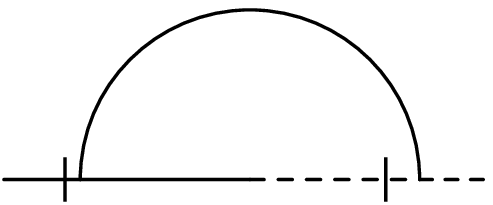}},
    \label{eq:vsphi}   \\
    \Gamma_{\phi' v} & = -iZ_5 p_i c^2 +\raisebox{-1ex}{ \includegraphics[width=2.5truecm]{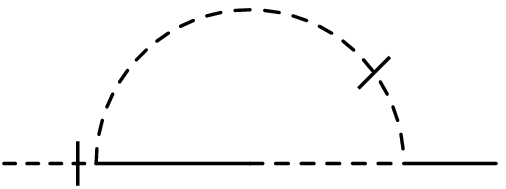}}
      +\raisebox{-1ex}{ \includegraphics[width=2.5truecm]{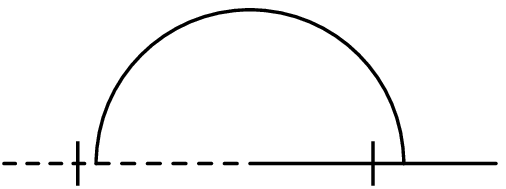}}
      +\raisebox{-1ex}{ \includegraphics[width=2.5truecm]{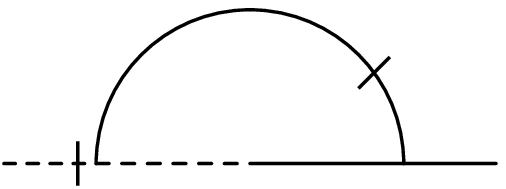}},
      \label{eq:phisv} \\
    \Gamma_{v'v'} & =  g_{1} \nu^3 p^{4-d-y} \biggl\{
  P_{ij}(\mp) + \alpha Q_{ij}(\mp)
  \biggl\} 
  + g_{2} \nu^3 \delta_{ij} Z_6 + \frac{1}{2}\raisebox{-1.0ex}{ \includegraphics[width=2.5truecm]{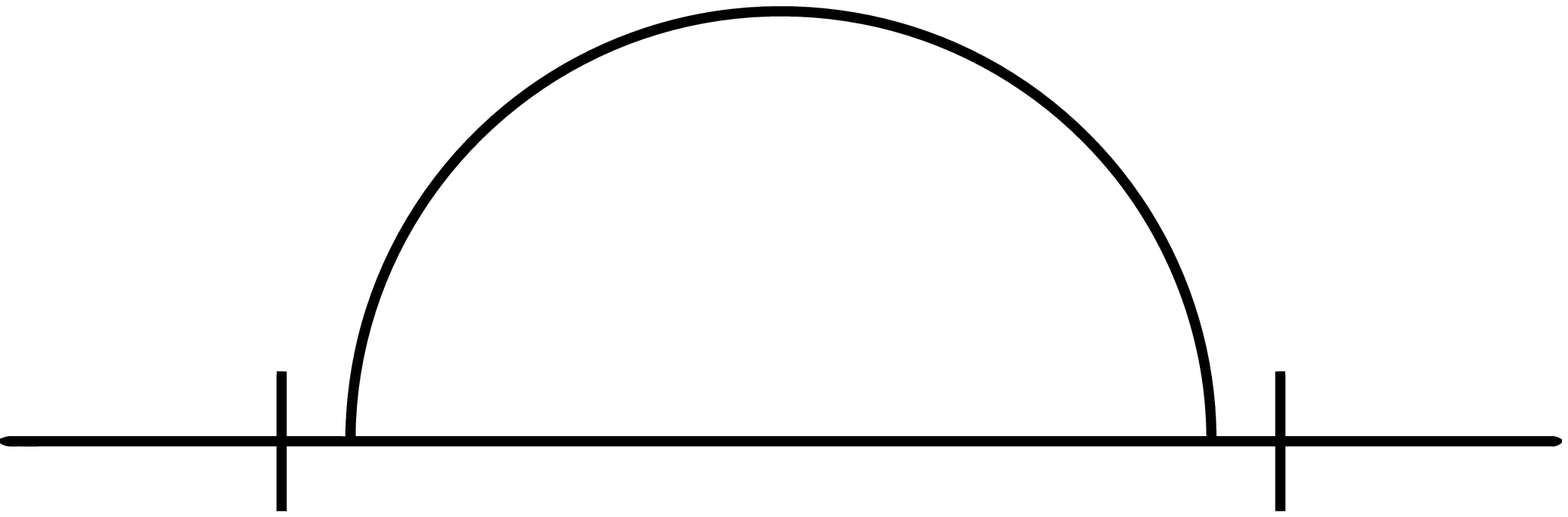}},
  \label{eq:vsvs}
\end{align}
where ${\bf p}$ always represents a corresponding external momentum. A factor $1/2$ in front of the diagram in~(\ref{eq:vsvs})
denotes a symmetry coefficient of the given graph.
Collecting all the mentioned facts and taking into account that non-local terms should not be renormalized,
it is straightforward to show that the theory is UV renormalizable. 
From the direct comparison of the relations between renormalized parameters it follows that
\begin{align}
   Z_\nu &= Z_1, &Z_{g_1}& = Z_1^{-3}, &Z_u& = Z_2 Z_1^{-1}, &Z_\phi& = Z_4, \nonumber\\
   Z_{\phi'} &= Z_4^{-1}, &Z_v& = Z_3 Z_{1}^{-1}, &Z_c& = (Z_4 Z_5)^{1/2}, &Z_{g_{2}}& = Z_6 Z_1^{-3}.
   \label{eq:RGconst2}
\end{align}
Employing dimensional regularization within minimal subtraction scheme (MS)~\cite{AGKL} the renormalization constants
can be calculated and the UV divergences manifests themselves in pole terms in $y$ and $\eps=4-d$. In higher loops
 pole terms in form of general linear combination in $ay+b\eps$ may appear.
{\section{UV renormalization of the model and fixed points} \label{sec:renorm}}
The large scale behavior with respect to spatial and time scales is
governed by the IR attractive stable fixed point $g^*\equiv\{g_1^*,g_2^*,u^*,v^*\}$. Here and henceforth 
the asterisk refers to a coordinate of the fixed point (FP).
Their coordinates are
determined from the relations~\cite{Vasiliev,Zinn} 
\begin{align}
  &\beta_{g_1} (g^{*}) = \beta_{g_2} (g^{*})= \beta_{u}
  (g^{*}) = \beta_{v} (g^{*}) = 0,
  \label{eq:gen_beta}
\end{align}
where $\beta_x=\widetilde{D}_\mu x$ for any variable $x$, and differential operator $\widetilde{D}_\mu $ denotes operation $\mu\partial_\mu$ at fixed bare parameters $\left\{g_{10}, g_{20}, u_0, v_0, \nu_0, \alpha_0, c_0\right\}$; $\mu$ is the ``reference mass'' (additional free parameter of the renormalized theory) in the MS renormalization scheme.
The eigenvalues of the matrix of first derivatives 
$\Omega_{ij}\equiv{\partial \beta_i}/{\partial g_j}$, where $i,j\in\{g_1,g_2,u,v\}$,
determine whether the given FP is IR stable or not. Points with all positive eigenvalues are
 candidates for macroscopic regimes and in principle
 can be observed experimentally. 
An explicit forms of the $\beta$-functions are
\begin{equation}
  \beta_{g_1} = g_1(-y - \gamma_{g_1}),\quad
  \beta_{g_2} = g_2(-\eps-\gamma_{g_2}),\quad
  \beta_u = -u\gamma_u,\quad
  \beta_v = -v\gamma_v,
  \label{eq:beta_functions}
\end{equation}
where $\gamma_x = \widetilde{D}_\mu \ln  Z_x$ are
the anomalous dimensions~\cite{Vasiliev}.
A direct analysis of the system of equations~(\ref{eq:gen_beta}) reveals the existence of three
IR stable fixed points: \fp{I}, \fp{II} and \fp{III}. \fp{I} is the free (Gaussian) fixed point, for 
which all interactions are irrelevant and no scaling and universality is expected. Its coordinates are	
\begin{equation}
  g_1^* = 0, \quad g_2^* = 0, \quad \text{whereas} \quad
  u^*\quad \text{and} \quad
  v^* \quad \text{are undetermined}.
  \label{eq:trivialFP}
\end{equation}
The corresponding eigenvalues of the matrix $\Omega$ are
\begin{equation}
   \lambda_1 = 0,\quad 
   \lambda_2 = 0,\quad
   \lambda_3 = -\eps,\quad 
   \lambda_4 = -y.
\end{equation}
Though trivial, this point is necessary for the correct use of perturbative renormalization group.

Further, there is a local fixed point \fp{II}, for which the charge $g_2$ attains a non-zero value, and corresponding
coordinates are
\begin{equation}
  g_1^* = 0, \quad g_2^* = \frac{8\eps}{3}, \quad
  u^* = 1, \quad
  v^* = 1.
\end{equation}
The eigenvalues of the matrix $\Omega$ are
\begin{equation}
   \lambda_1 = \frac{7\eps}{18},\quad 
   \lambda_2 = \frac{5\eps}{6},\quad
   \lambda_3 = \eps,\quad 
   \lambda_4 = \frac{3\eps-2y}{2}.
\end{equation}
For the last fixed point, \fp{III}, both non-local and local parts of the random force are relevant:
\begin{equation}
  g_1^* = \frac{16y(2y-3\eps)}{9(y(2+\alpha)-3\eps)}, 
  \quad g_2^* =\frac{16\alpha y^2}{9(y(2+\alpha)-3\eps)}
, \quad
  u^* = 1, \quad
  v^* = 1;
\end{equation}
the required eigenvalues are
\begin{equation}
   \lambda_1 = \frac{y[2y(10\alpha+11)-3\eps(3\alpha+11)]}{54[y(2+\alpha)-3\eps]} ,\quad
   \lambda_2 =\frac{y[2y(2\alpha+3)-\eps(\alpha+9)]}{6[y(\alpha+2)-3\eps]}, \quad
   \lambda_{3,4} = \frac{A\pm \sqrt{B}}{C},   
\end{equation}
where $A,B$ and $C$ are given by the following expressions:
\begin{align}
  A & = -27 \eps^3+9 (9+\alpha) \eps^2 y-9 (8+3 \alpha)    \eps y^2+
  2y^3(\alpha^2+7\alpha+10);\\
  B & =[-3 \eps + (2+\alpha) y]^2 [81 \eps^4-54 \eps^3 y-9 (3+20 \alpha) \eps^2 y^2+12
  (1+17 \alpha+3 \alpha^2) \eps y^3 \nonumber\\
  &-4 (-1+14 \alpha+5 \alpha^2) y^4];\\
  C & = 6 [-3 \eps+(2+\alpha) y]^2.
\end{align} 

From the physical interpretation of the kernel function~(\ref{eq:correl}) it follows that the charges
$g_1^*$ and $g_2^*$ can not attain negative values. Using this fact together with an explicit form of the eigenvalues $\lambda_1\ldots\lambda_4$
it can be shown, that the point \fp{III} is stable for $y>0$ and $y>3\eps/2$. 
Note, that the crossover between two nontrivial points happens along the line $y=3\eps/2$, which is in accordance with~\cite{Ant04}.
{\section{Conclusion} \label{sec:consl}}
In this paper the compressible extension of the stochastic Navier Stokes equation
has been studied using the field theoretic approach.
Crucial points of the Feynman diagrammatic technique and perturbative renormalization group
have been discussed. One loop approximation provides that, depending of the exponent $y$ and
deviation from the dimension of ${\mx}$ space $\eps=4-d$, the model possesses three stable 
fixed points in the IR region (i.e., three possible scaling regimes)~-- trivial (Gaussian, \fp{I}), local (\fp{II}) and nonlocal (\fp{III}).

This shows us, that the simple analysis around $d=3$, which indicates existence of only one
nontrivial fixed point~\cite{ANU97}, is incomplete in this case.



\begin{acknowledgement}

The authors are indebted to M.~Yu.~Nalimov, L.~Ts.~Adzhemyan, M.~Hnati\v{c}, J.~Honkonen, and V.~\v{S}kult\'ety for discussions.

The work was supported by the Saint Petersburg State University within the
research grant 11.38.185.2014 
and by the Russian Foundation for Basic
Research within the Project 16-32-00086.
N.~M.~G. and M.~M.~K. were also supported by the Dmitry Zimin's ``Dynasty'' foundation.

\end{acknowledgement}


\end{document}